\documentclass[%
 reprint,
 amsmath,amssymb,
 aps,
]{revtex4-2}

\usepackage{graphicx}
\usepackage{dcolumn}
\usepackage{bm}
\usepackage{xcolor}

\begin{document}

\preprint{APS/123-QED}



\title{Experimental Evidence for Increased Particle Fluxes Due to a Change in Transport at the Separatrix near Density Limits on Alcator C-Mod}

\author{M.A. Miller$^{1}$, J.W. Hughes$^{1}$, T. Eich$^{2}$, G.R. Tynan$^{3}$, P. Manz$^{4}$, A.E. Hubbard$^{1}$, B. LaBombard$^{1}$, J. Dunsmore$^{1}$}

\address{$^1$MIT Plasma Science and Fusion Center, Cambridge, MA 02139, USA
}
\address{$^2$Commonwealth Fusion Systems, Devens, MA 01434, USA
}
\address{$^3$University of California, San Diego, CA 92093, USA
}
\address{$^4$Institute of Physics, University of Greifswald, Felix-Hausdorff-Str.6, Greifswald, 17489, Germany
}

\date{\today}

\begin{abstract}
Experimental inferences of cross-field particle flux at the separatrix, $\Gamma_{\perp}^\mathrm{sep}$, show rapid growth near H-mode and L-mode density limits at high magnetic field on Alcator C-Mod. Increases in $\Gamma_{\perp}^\mathrm{sep}$ correlate well with proximity to high density operational boundaries as proposed by the separatrix operational space model. $\Gamma_{\perp}^\mathrm{sep}$ grows as the L-mode density limit and the H-L-mode back transition boundaries are approached, consistent with expectations of plasma instability-driven turbulence suggested by theory, confirming the power dependence of density limits. $\Gamma_{\perp}^\mathrm{sep}$ is well-organized by the characteristic wavenumber for resistive ballooning mode turbulence, $k_\mathrm{RBM}$, from interchange-drift-Alfvén fluid turbulence theory, with additional dependence on the cylindrical safety factor, $\hat{q}_\mathrm{cyl}$, yielding an empirical limit to plasma operation of $k_\mathrm{RBM}^{2}\hat{q}_\mathrm{cyl} = 1$. This limit corresponds to the point where the perpendicular heat flux, $Q_{\perp}$, reaches the level of the parallel heat flux, $Q_{\parallel}$, i.e. $Q_{\perp} \approx Q_{\parallel}$, beyond which point thermal equilibrium is not satisfied, resulting in a fold catastrophe.
\end{abstract}

\maketitle

There is a strong need to develop a physics-based understanding of mechanisms that limit the edge density profile of tokamaks operating at high density. Fusion reactors will need to operate with high edge densities \cite{kallenbach_analytical_2016, goldston_new_2017, Moulton_Lengyel_2021, eich_separatrix_2025} to ensure that power fluxes to the device remain below material limits \cite{gunn_surface_2017, reinke_heat_2017, kuang_divertor_2020, menard_fusion_2022}. Most importantly, operation at high density puts plasmas at risk of density-limit-driven disruptions \cite{greenwald_new_1988, petrie_plasma_1993}, the triggering physical mechanism of which remains contested. A growing body of evidence exists suggesting that at high density, the role of plasma transport will eclipse that of ionizing neutrals in determining the edge density \cite{labombard_particle_2001, hughes_edge_2007, mordijck_impact_2024}. In these reactor-relevant scenarios, use of external actuators to modify the plasma density may be limited. It is therefore imperative to understand particle transport at high density, its implications for core performance, and crucially, its influence in bringing a plasma toward high density operational limits.



Using novel experimental analysis, this Letter finds that particle transport plays a key role in the limit to plasma operation at high density on the Alcator C-Mod tokamak. Unique diagnostic capability on C-Mod of neutral Ly$_{\alpha}$ emission in the edge enabled inference of the cross-field particle flux at the separatrix, $\Gamma_{\perp}^\mathrm{sep}$. Inferences of $\Gamma_{\perp}^\mathrm{sep}$ are collated into databases and compared against key parameters describing drift-Alfvén turbulence (DALF) \cite{scott_three-dimensional_1997, scott_computation_1998, scott_computation_2003, scott_energetics_2005}, codified by boundaries from the recently proposed separatrix operational space (SepOS) model \cite{eich_separatrix_2021}. Large increases in $\Gamma_{\perp}^\mathrm{sep}$ are found at high-density turbulent transition boundaries from the SepOS model. These changes track $\alpha_{t}$, a parameter mediating the influence of interchange modes over drift-wave (DW) modes \cite{eich_turbulence_2020}. Across both the low-confinement mode (L-mode) and the high-confinement mode (H-mode), $\Gamma_{\perp}^\mathrm{sep}$ organizes best with the normalized characteristic wavenumber for the resistive ballooning mode (RBM), $k_\mathrm{RBM}$, derived from the SepOS model. The data suggest that particle transport prior to the H- and L-mode density limits has a common physical origin rooted in the interchange-driven RBM instability \cite{bateman_resistive-ballooning-mode_1978, carreras_transport_1983, mccarthy_stability_1992}. Limits to operation in both regimes and across a range of plasma current, $I_{P}$, are well-described by an approach to a critical value $k_\mathrm{RBM}^{2}\hat{q}_\mathrm{cyl} = 1$, with $\hat{q}$ the cylindrical safety factor.

Under reasonable assumptions, this empirical limit can be expressed as a scaling for the pressure gradient scale length, $\lambda_{p}$, of the form $\lambda_{p} \sim a$, a key assumption in the density limit scaling proposed in the recent work by Giacomin et al. \cite{giacomin_first-principles_2022}. The limit also suggests that the transport scales stronger-than-linearly with $\alpha_{t}$, consistent with recent experimental observations \cite{eich_turbulence_2020, li_study_2025}. Further, this empirical limit yields a scaling on power, $P$ for the density limit of the form $n \sim P^{4/7}$, similar to that from a model for radiative collapse \cite{stroth_model_2022}. The primary scaling on power of the threshold limit suggests that ultimate thermal collapse is driven by atomic physics. Prior to this limit, large particle fluxes drive strong perpendicular heat fluxes, $Q_{\perp}$, on the order of the parallel heat flux, $Q_{\parallel}$, directly preceding thermal catastrophe as proposed in a theory for the density limit related to thermal instability \cite{dippolito_thermal_2006}. The transport analysis shown in this Letter experimentally affirms the role of turbulence in reaching thermal catastrophe via power balance, in line with earlier conjectures \cite{labombard_particle_2001, greenwald_density_2002, manz_power_2023, manz_density_2025} and bridging the gap between disparate paradigms for which to understand the density limit.

\begin{figure*}
\centering
\includegraphics[width=2\columnwidth]{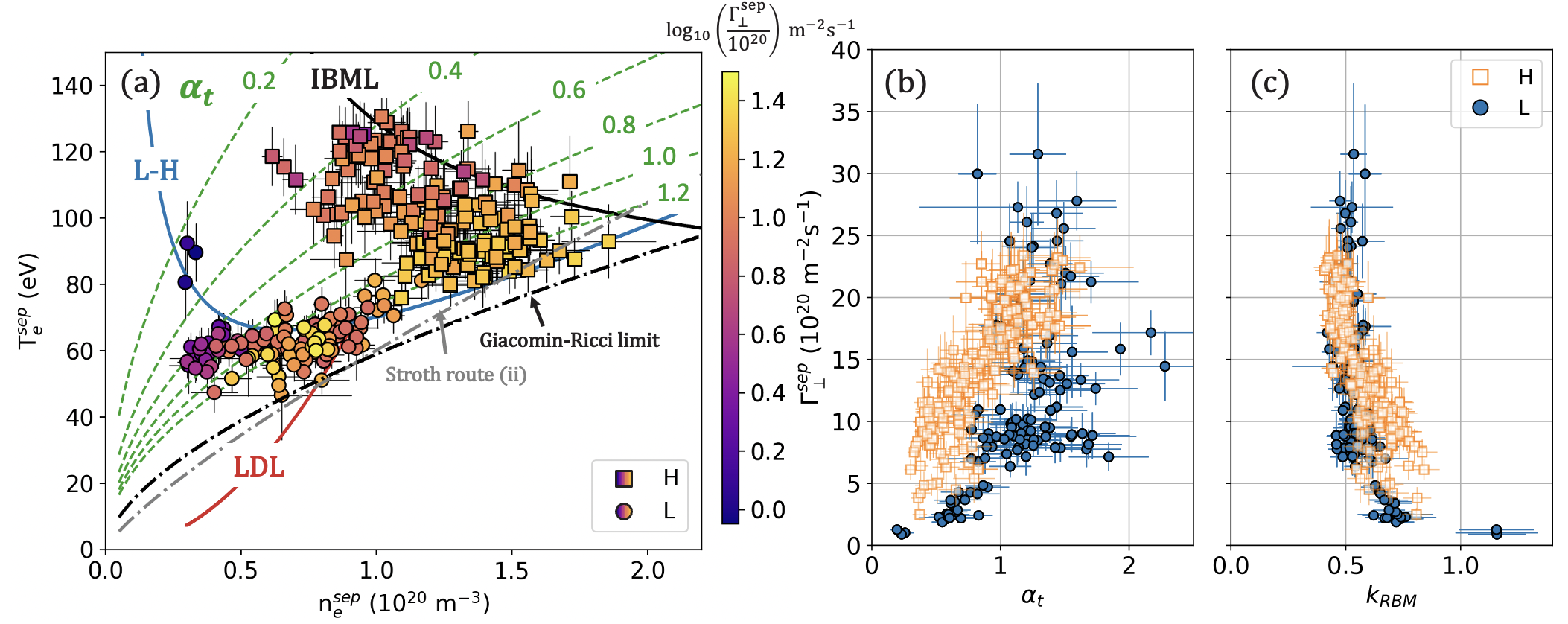}
\caption{(a) Separatrix operational space for subset of discharges at fixed $I_{P}$, $B_{t}$, and shape, all containing Ly$_{\alpha}$. The color variable shows $\mathrm{log}(\Gamma_{\perp}^\mathrm{sep})$. (b) $\Gamma_{\perp}^\mathrm{sep}$ plotted against $\alpha_{t}$ and (c) against $k_\mathrm{RBM}$. H-modes are shown as squares and L-modes are shown as circles.}
\label{fig:fluxes_fixed_ip}
\end{figure*}



On Alcator C-Mod, $\Gamma_{\perp}$ is obtained experimentally for many discharges through measurement of the electron density, $n_{e}$, and electron temperature, $T_{e}$, as well as the ionization source via measurements of line-integrated Ly$_{\alpha}$ emissivity, $\epsilon_{\mathrm{Ly}_{\alpha}}$, from the population of neutral atoms in the edge plasma. A collisional-radiative model, ADAS \cite{adas}, allows inference of $\Gamma_{\perp}$, under the assumption of insignificant poloidal asymmetry and the imposition of stationarity (ensured by filtering the time rate of change of the stored energy, $dW/dt < 0.1$ MJs$^{-1}$). Details of the computation can be found in \cite{miller_enhanced_2025} and \cite{miller_particle_2025}.



Using these measurements, this Letter builds on recent work that confirmed that the separatrix operational space of C-Mod is well-described by the boundaries proposed by the SepOS model \cite{miller_determination_2025}. The model was originally developed using data from ASDEX Upgrade (AUG) \cite{eich_separatrix_2021} and the equations of DALF turbulence theory \cite{scott_three-dimensional_1997, scott_computation_1998} and then applied to a C-Mod database without Ly$_{\alpha}$ measurements \cite{miller_determination_2025}. The SepOS model describes three main boundaries in the tokamak operational space at the separatrix: the transition from L-mode to H-mode (L–H), the L-mode density limit (LDL), and the ideal ballooning magnetohydrodynamic (MHD) limit (IBML). The model can be recast into an intuitive formalism suited for fusion device design and operation in boundaries parameterized by $n_{e}$ and $T_{e}$, as shown in Figure \ref{fig:fluxes_fixed_ip}(a). The availability of Ly$_{\alpha}$ measurements in this dataset and resulting inferences of $\Gamma_{\perp}$ presented in this Letter allow for direct testing of the physical mechanisms behind the boundaries of the separatrix operational space.

Figure \ref{fig:fluxes_fixed_ip}(a) shows the operational space in terms of separatrix density, $n_{e}^\mathrm{sep}$, and separatrix temperature, $T_{e}^\mathrm{sep}$, for a subset of plasmas at fixed plasma current, $I_{P} = 0.8 \pm 0.1$ MA, toroidal magnetic field, $B_{t} = 5.4 \pm 0.1$ T, and relatively fixed shape, with elongation $\kappa = 1.65 \pm 0.05$, distance between primary and secondary separatrix, $\Delta R_\mathrm{sep} < 0$ mm, i.e. all in lower single null, and with some range in triangularity, $\delta = \frac{\delta_{l} + \delta_{u}}{2}$ between 0.36 -- 0.54. H-modes are shown as squares and L-modes are shown as circles. Separatrix quantities are estimated from power balance in the scrape-off layer (SOL) using the two-point model \cite{stangeby_pc_plasma_2000}. Details of the procedure for estimating these quantities can be found in \cite{miller_determination_2025}. The three main SepOS boundaries described above are computed and also plotted in Figure \ref{fig:fluxes_fixed_ip}(a), with the L-H in blue, the LDL in red, and the IBML in black. This figure extends the recent validation effort model \cite{miller_determination_2025} showing good agreement between C-Mod separatrix quantities and the boundaries from the SepOS to a slightly larger dataset, but crucially, one also including $\Gamma_{\perp}$, shown using the color variable. It is evident that $\Gamma_{\perp}^\mathrm{sep}$ increases as L-modes approach the LDL or H-modes approach the high-density portion of the L-H transition curve (equivalently, the H-L back transition curve). These have both been previously proposed as disruptive limits to high-density operation for L- and H-mode \cite{manz_power_2023, manz_how_2025, manz_density_2025}. Notably, H-modes approaching the IBML boundary and L-modes approaching the L-H boundary do not experience systematic increases in $\Gamma_{\perp}^\mathrm{sep}$.

Changes to $\Gamma_{\perp}^\mathrm{sep}$ are aptly represented with $\alpha_{t}$, the contours of which are shown as green dashed lines. $\alpha_{t}$ captures the cross-phase of potential and pressure fluctuations and is defined by $\alpha_{t} \equiv C\omega_{B}$, where $C$ is a normalized collision frequency, introduced in \cite{scott_three-dimensional_1997} and $\omega_{B} = \frac{2\lambda_{p}}{R}$ gives the strength of the curvature drive \cite{eich_separatrix_2021}, with $R$ the major radius. The product $C\omega_{B}$ scales with $\hat{q}_\mathrm{cyl}\nu^{*}$, with $\nu^{*}$ the collisionality. Though the definition of $\alpha_{t}$ includes a drive term, $\omega_{B}$, it is ultimately canceled out by that in $C$, such that the product in practice yields contours of constant $\nu^{*}$ when plotted in ($n_{e}$, $T_{e}$) space (at fixed $\hat{q}_\mathrm{cyl}$). $\alpha_{t}$ is related to electron adiabaticity, a quantity previously linked to changes to particle transport near density limits \cite{hong_edge_2018, long_enhanced_2021} on HL-2A and J-TEXT. On both devices near these limits, decreased adiabaticity was tied to increased turbulent particle transport and decreased Reynolds power, the mechanism thought to transfer energy from turbulence to mean flows \cite{diamond_self-regulating_1994, kim_zonal_2003, estrada_experimental_nodate, manz_zonal_2012, tynan_turbulent-driven_2013}. They show shear layer collapse as the LDL is approached, further experimental evidence of the key role of edge turbulent transport on approach to the density limit. These measurements are consistent with theory suggesting that Reynolds energy transfer acts to lower turbulent energy and prevents collapse of the shear layer associated with the LDL \cite{greenwald_density_2002, diamond_how_2023}. Here, for fixed $\alpha_{t}$, H-modes have \emph{larger} $\Gamma_{\perp}$ than L-modes, despite their improved confinement. Some mechanism must then be responsible for sustaining the large mean flow of H-modes even in the case of larger transport -- energy transfer via the Reynolds stress is a good candidate. At high $\alpha_{t}$, Reynolds energy transfer efficiency decreases, and the energy associated with large $\Gamma_{\perp}$ can no longer be effectively transferred into the shear flow, leading to a density limit. H-modes in the corner between the H-L and IBML curves have high $\Gamma_{\perp}$ not as a result of proximity to the IBML but rather the H-L curve, which ultimately sets particle transport prior to the density limit.

\begin{figure}
\centering
\includegraphics[width=\columnwidth]{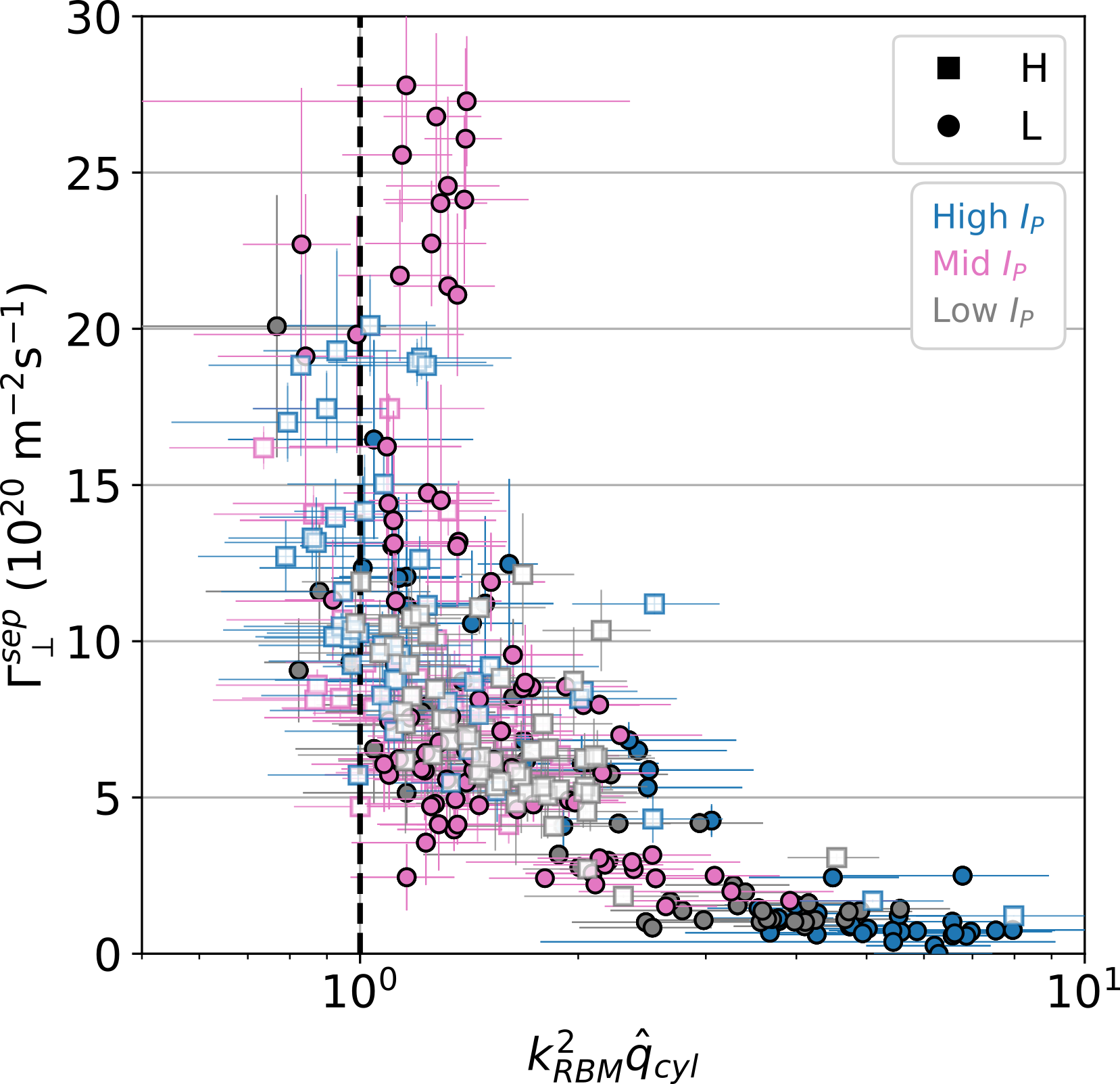}
\caption{$\Gamma_{\perp}^\mathrm{sep}$ plotted against $k_\mathrm{RBM}^{2}\hat{q}_\mathrm{cyl}$ for low $I_{P}$ (gray), mid $I_{P}$ (pink), and high $I_{P}$ (mid) discharges in H-mode (squares) or L-mode (circles). The vertical black dashed line represents $k_\mathrm{RBM}^{2}\hat{q}_\mathrm{cyl}=1$.}
\label{fig:gamma_krbm_qcyl}
\end{figure}



A scaling for the density limit has been recently proposed based on similar arguments of increased particle transport \cite{giacomin_first-principles_2022}. The limit was derived by relating higher turbulent transport with the local flattening of the pressure gradient, eventually leading to disruption. The Giacomin-Ricci limit can be expressed in terms of $n_{e}^\mathrm{sep}$ and $T_{e}^\mathrm{sep}$ using the same experimental methodology as in \cite{manz_power_2023}, yielding the dash-dotted black curve in Figure \ref{fig:fluxes_fixed_ip}(a). Indeed, this curve acts as a limit to all data at high $n_{e}$ and low $T_{e}$. It aligns well with discharges with high $\Gamma_{\perp}$, consistent with elevated turbulent transport proposed by the model. Proximity to disruptive limits can also be understood through power balance by comparison to the Stroth model for an unstable X-point MARFE \cite{stroth_model_2022}. Route (ii) of this model suggests that this can occur if an X-point radiator cools the plasma near the edge enough to yield enhanced transport via resistive MHD. This occurs at a critical $n_{e}$, which varies with $T_{e}$, shown as a dash-dotted gray line in Figure \ref{fig:fluxes_fixed_ip}(a). As discharges approach this limit, fluxes increase, decreasing the amount of power available per particle, and cooling the plasma below sustainable limits. This leads to termination of the H-mode or of the plasma entirely if the L-mode cannot be sustained with the remaining power. Ideas about the role of power balance presented by Zanca \cite{zanca_power-balance_2019} were found to be consistent with those from route (ii) of the Stroth model \cite{manz_power_2023}.





To query the governing physics further, Figures \ref{fig:fluxes_fixed_ip}(b) and (c) plot $\Gamma_{\perp}$ against local turbulence control parameters. Figure \ref{fig:fluxes_fixed_ip}(b) shows that $\Gamma_{\perp}$ grows quickly as $\alpha_{t}$ reaches large values ($>1$), at which point interchange modes are thought to dominate over DW modes \cite{eich_turbulence_2020}. Figure \ref{fig:fluxes_fixed_ip}(c) plots $\Gamma_{\perp}$ against $k_\mathrm{RBM}$, introduced in \cite{eich_separatrix_2021} and defined as $k_\mathrm{RBM} \equiv \sqrt{\frac{\alpha_{c}\sqrt{\omega_{B}}}{\alpha_{t}}}$, with $\alpha_{c}$ the critical normalized pressure gradient and $\omega_{B}$. $k_\mathrm{RBM}$ is similar to the diamagnetic parameter, $\alpha_{d}$, from seminal work on edge turbulence-dictated phase spaces \cite{rogers_phase_1998}. Generally, $k_\mathrm{RBM}$ offers better cross-regime organization of $\Gamma_{\perp}$ than $\alpha_{t}$. Unlike $\alpha_{t}$, $k_\mathrm{RBM}$ also depends on $\lambda_{p_{e}}$ through $\omega_{B}$, which normalizes the turbulence driving parameter across regimes. The exact value of $\alpha_{t}$ at which large $\Gamma_{\perp}$ exists varies between H- and L-modes, with L-modes accessing larger values of $\alpha_{t}$, whereas large growth in $\Gamma_{\perp}$ occurs for $k_\mathrm{RBM} \lesssim 0.6$ for both L- and H-modes. Additionally, all discharges are limited by $k_\mathrm{RBM} \approx 0.4$, corresponding to the the LDL and the H-L boundaries. On AUG, discharges are also limited by the same value \cite{eich_separatrix_2021}. Unnormalizing the empirical limit in $k_\mathrm{RBM}$ gives a characteristic fluctuation wavelength, $\lambda = \frac{2\pi \rho_{s}}{k_\mathrm{RBM}} \approx 15 \rho_{s} \approx \rho_{s,p}$, indicating that a density limit may be reached when the typical wavelength of RBM fluctuations reaches the size of the poloidal gyroradius, $\rho_{s,p}$.

\begin{figure*}
\centering
\includegraphics[width=1.7\columnwidth]{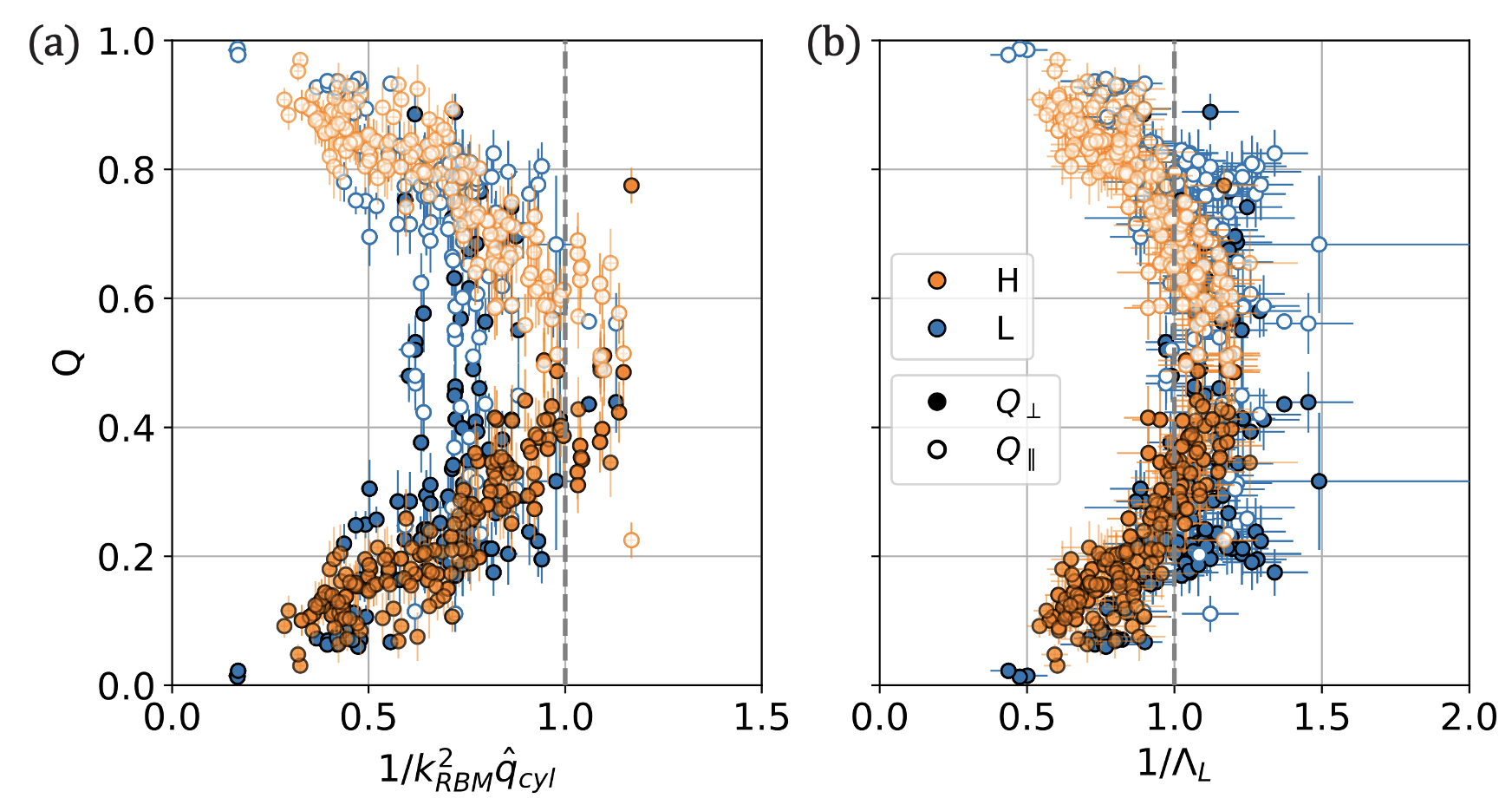}
\caption{Fraction of heat flux in perpendicular (closed circles) and parallel directions (open circles) against (a) $1/k_\mathrm{RBM}^{2}\hat{q}_\mathrm{cyl}$ and (b) $1/\Lambda_{L}$ for H-modes (orange) and L-modes (blue). Vertical lines highlight unity on the abscissa.}
\label{fig:heat_flow}
\end{figure*}

Since this first dataset is at fixed $I_{P}$ (and $B_{t}$), or equivalently, fixed $\hat{q}_\mathrm{cyl}$, a second dataset is used to investigate the impact of $I_{P}$ variation on $\Gamma_{\perp}$ near disruptive boundaries. This dataset features variation of up to double in $I_{P}$, from 0.6 -- 1.2 MA, at fixed $B_{T} \approx 5.4$ T. Discharges are grouped into categories at low $I_{P} = 0.65 - 0.75$ MA in blue, mid $I_{P} = 0.75 - 0.85$ MA in pink, and high $I_{P} = 0.85 - 1.25$ MA in gray, with most discharges below 1 MA. In Figure \ref{fig:gamma_krbm_qcyl}, $\Gamma_{\perp}^\mathrm{sep}$ is plotted against the square of $k_\mathrm{RBM}$, multiplied by $\hat{q}_\mathrm{cyl}$, which accounts for variation in $I_{P}$. For a given $k_\mathrm{RBM}$, discharges at high $I_{P}$ have higher $\Gamma_{\perp}$ than those at low $I_{P}$. The inclusion of $\hat{q}_\mathrm{cyl}$ on the abscissa accounts for this by shifting low-$I_{P}$ (high-$\hat{q}_\mathrm{cyl}$) discharges to the right and high-$I_{P}$ (low-$\hat{q}_\mathrm{cyl}$) discharges to the left. Plotting against the parameter $k^{2}_\mathrm{RBM}\hat{q}_\mathrm{cyl}$ reduces scatter and provides the best cross-regime organization of $\Gamma_{\perp}^\mathrm{sep}$.

This figure shows that the large increases in $\Gamma_{\perp}^\mathrm{sep}$ observed near disruptive limits in both H- and L-modes correspond to an approach of $k_\mathrm{RBM}^{2}\hat{q}_\mathrm{cyl} \rightarrow 1$. The empirical limit $k_\mathrm{RBM}^{2}\hat{q}_\mathrm{cyl} = 1 $ is best explained by contact with a theory for thermal collapse based on blob-dominated convective transport \cite{dippolito_thermal_2006}. There, the density limit is classified as a fold catastrophe resulting from root merging. Thermal instability occurs when the physical and unphysical solutions of $Q_{\perp}$ and $Q_{\parallel}$ merge, which occurs just after $Q_{\perp} > Q_{\parallel}$ for the physical roots (see Figure 3 in \cite{dippolito_thermal_2006}). Experimentally, $Q_{\parallel}$ can be readily estimated using the Spitzer-Härm assumption for parallel transport, i.e $Q_{\parallel} = \kappa_{0,e}T_{e}^{5/2}\nabla_{\parallel}T_{e}S_{\parallel} \approx \kappa_{0,e}\frac{T_{e}^{7/2}}{L_{\parallel}}S_{\parallel}$, with $\kappa_{0,e}$ the parallel electron conductivity coefficient,
$L_{\parallel}$ the connection length to the outer target, and $S_{\parallel}$ the area of a flux tube just outside the separatrix of width $\lambda_{q}$, the parallel heat flux decay length. $Q_{\perp}$ can be estimated experimentally using $\Gamma_{\perp}$ under the assumption that $Q_{\perp}^\mathrm{conv} \gg Q_{\perp}^\mathrm{cond}$, using $Q_{\perp} = \frac{3}{2}T_{e}\Gamma_{\perp}S_{\perp}$, with $S_{\perp}$ the surface area of the last closed flux surface. The assumption of convective-dominated transport is generally reasonable near the separatrix, and especially so near density limits, as is also assumed in \cite{giacomin_first-principles_2022} and \cite{stroth_model_2022}.

Figure \ref{fig:heat_flow} plots these two quantities normalized to their sum, $Q_{\perp} + Q_{\parallel}$, for the fixed $I_{P}$ dataset. Figure \ref{fig:heat_flow}(a) plots $Q$ against the inverse of $k_\mathrm{RBM}^{2}\hat{q}_\mathrm{cyl}$ and Figure \ref{fig:heat_flow}(b) plots it against the inverse of $\Lambda_{L}$, defined as $\Lambda_{L} \equiv \frac{1}{q}\left(\frac{\lambda_{ei}}{R}\right)^{1/2}$, with $\lambda_{ei}$ the electron-ion collision mean free path. $\Lambda_{L}$ was proposed as an inverse collisionality parameter associated with the approach to critical gradients in ohmic plasmas on C-Mod \cite{labombard_critical_2008}. Note that this and its preceding works \cite{labombard_particle_2001, labombard_evidence_2005} provided the experimental foundations for the development of the thermal catastrophe theory \cite{dippolito_thermal_2006} and the groundwork on which a large part of analysis of this Letter is built. Figure \ref{fig:heat_flow} shows notable similarities with the physical model solutions presented in Figure 3 of \cite{dippolito_thermal_2006}, noting that $k_\mathrm{RBM}^{2}\hat{q}_\mathrm{cyl}$ and $\Lambda_{L}$ both scale inversely with collisionality. As the inverse of these parameters increases, $Q_{\perp}$ increases, while $Q_{\parallel}$ decreases, in both H-mode and L-mode, until $Q_{\parallel} \approx Q_{\perp} \approx \frac{1}{2}Q$. Beyond a critical value of approximately unity for $1/k_\mathrm{RBM}^{2}\hat{q}_\mathrm{cyl}$ and $1/\Lambda_{L}$, data are scarce, indicative of instability. As $1/\Lambda_{L}$ approaches and crosses this value, the blob perpendicular velocity increases and plasmas transition into the ``disconnected blob'' regime of \cite{dippolito_thermal_2006}. It is only when $1/k_\mathrm{RBM}^{2}\hat{q}_\mathrm{cyl} \rightarrow 1$ that a final stability limit on thermal equilibrium is reached. After $k_\mathrm{RBM}^{2}\hat{q}_\mathrm{cyl} > 1$, $Q_{\perp}$ reaches and begins to surpass $Q_{\parallel}$, beyond which their physical and unphysical roots merge and plasmas cannot remain stable.

The limit $k_\mathrm{RBM}^{2}\hat{q}_\mathrm{cyl} = 1$ contains scalings providing a link to earlier experimental and theoretical work. Expressing it in terms of its constituent components yields $\lambda_{p_{e}} = \frac{\alpha_{t}^{2}R}{\alpha_{c}^{2}\hat{q}_\mathrm{cyl}^{2}}$, which matches the H-mode scaling dependence of $\lambda_{p_{e}}$ on $\alpha_{t}$ found on AUG \cite{eich_turbulence_2020}. Similar stronger-than-linear dependence of edge gradient scale lengths on $\alpha_{t}$ has been also recently observed in high-density H-modes on EAST \cite{li_study_2025}. Near density limits, $\alpha_{t}$ reaches a fixed, maximum value, and for a fixed geometry, the scaling shown here gives $\lambda_{p_{e}} \sim R$. For fixed aspect ratio one recovers $\lambda_{p_{e}} \sim a$, providing an explanation for a key assumption in the Giacomin-Ricci model \cite{giacomin_first-principles_2022}. Despite including wavenumbers for RBM turbulence, the empirical limit found here points to a physical phenomenon related to atomic and radiation physics. One can use the assumption $T_{e} \sim \left(\frac{P_\mathrm{sep} \hat{q}_\mathrm{cyl}^{2}}{\lambda_{q}}\right)^{2/7}$ and solve for $n_{e} \sim \frac{\alpha_{c} \hat{q}_\mathrm{cyl}^{1/7}}{Z_\mathrm{eff}R_\mathrm{geo}^{3/2}} \frac{\lambda_{p}^{1/2}}{\lambda_{p}^{4/7}} P_\mathrm{sep}^{4/7}$. Ignoring the small $\hat{q}_\mathrm{cyl}$ dependence, taking $\lambda_{p} \sim \lambda_{q}$, and ignoring the remaining small gradient scale length dependence, one arrives at the density scaling, $n_{e} \sim \frac{\alpha_{c}}{Z_\mathrm{eff}R_\mathrm{geo}^{3/2}}P_\mathrm{sep}^{4/7}$. For a fixed shape plasma, the limit then primarily depends on $P_\mathrm{sep}$, the net power through the separatrix, with an exponent similar to that of recent works \cite{stroth_model_2022, giacomin_first-principles_2022}. For the ITER baseline scenario ($I_{P} = 15$ MA, $a = 2.0$ m, $P_\mathrm{sep} = 50$ MW) \cite{loarte_new_2025}, this corresponds to a limiting density, $n_{e}^\mathrm{sep} = 1.1 \times 10^{20}$ m$^{-3}$, just below the Greenwald density, $n_{G}$ for that scenario, $n_{G} = 1.2 \times 10^{20}$ m$^{-3}$. Even for a flat density profile, with $n_{e}^\mathrm{sep}/\overline{n}_{e} \approx 0.5$, this represents nearly a doubling of the predicted limiting $n_{e}$. Repeating the exercise for SPARC's primary reference discharge ($I_{P} = 8.7$ MA, $a = 0.57$ m, $P_\mathrm{sep} = 28$ MW) \cite{creely_overview_2020} yields $n_{e}^\mathrm{sep} = 4.7 \times 10^{20}$ m$^{-3}$. Assuming once more $n_{e}^\mathrm{sep}/\overline{n}_{e} \approx 0.5$ gives a prediction close to $n_{G} = 8.5 \times 10^{20}$ m$^{-3}$. These predictions closely match those made for both the ITER and SPARC scenarios in \cite{giacomin_first-principles_2022}, demonstrating the importance of considering the power dependence of this limit.

The unfolding of this limit suggests that the end chain of events is linked to radiation and thus ultimately determined by atomic physics. But, that $\Gamma_{\perp}$ organizes against parameters associated with the RBM for high density plasmas on Alcator C-Mod indicates, however, that the RBM (or an instability that scales like it) plays a large role in driving transport through the separatrix. Similarly to $\Lambda_{L}$, the importance of $k_\mathrm{RBM}$ in regulating transport may also explain insensitivity of edge density profiles observed at high $n_{e}$ often observed on Alcator C-Mod in \emph{both} L- and H-mode \cite{hughes_edge_2007, labombard_critical_2008, miller_enhanced_2025} -- resistive modes drive strong transport, placing a limit on the density gradient of high density profiles. If $1/\Lambda_{L} > 1$ describes the transition to a convective, blob-dominated turbulence regime where such an instability is active, the limit $1/k_\mathrm{RBM}^{2}\hat{q}_\mathrm{cyl} \rightarrow 1$ is ultimately responsible for thermal catastrophe and collapse of the plasma column.

This Letter provides experimental evidence for the impact of turbulent cross-field particle transport in setting operational limits at high density in both H- and L-modes on Alcator C-Mod. It has clarified the local physics parameter determining the cross-field particle flux, acting as a powerful extension of a large body of work studying the role of dimensionless edge turbulence control parameters in limiting plasma operation \cite{diamond_self-regulating_1994, rogers_enhancement_1997, scott_three-dimensional_1997, kim_zonal_2003, labombard_evidence_2005, fedorczak_shear-induced_2012, tynan_turbulent-driven_2013, cziegler_fluctuating_2013}. The new findings presented here are enabled by neutral measurements allowing experimental inference of the particle flux. The presence of both ETS and LYMID enables particle balance calculations and the comparison of particle transport data with parameters from edge turbulence and stability theory. The empirical limit found in this work suggest that the final step in the chain of events associated with the density limit is the collapse of a power-starved plasma via atomic physics. Scalings of $\Gamma_{\perp}^\mathrm{sep}$ and $Q_{\perp}$, however, confirm that it is enhanced perpendicular transport \emph{across} relative to parallel transport \emph{along} the separatrix, that is fundamentally responsible for the edge plasma cooling leading to the density limit. Power-dependent transport phenomena preceding high density limits and their impact on edge profiles may well be important for next-step devices and power plants like SPARC \cite{creely_overview_2020, eich_separatrix_2025} and ARC \cite{sorbom_arc_2015}, which plan to operate at higher densities and magnetic fields, and for which understanding limits to edge density profiles will be of utmost importance for sustainable and efficient operation. 





\bibliography{citations}

\end{document}